\documentclass[%
 reprint,
 superscriptaddress,
 amsmath,amssymb,
 aps,longbibliography
]{revtex4-1}

\usepackage{graphicx}
\usepackage{dcolumn}
\usepackage{bm}
\usepackage{hyperref}
\usepackage[mathlines]{lineno}
\usepackage[bottom]{footmisc}
\usepackage[super]{nth}%
\usepackage{color}
\usepackage{hhline}

\begin{document}
\raggedbottom


\title{Subdiffusive light transport in three-dimensional subrandom arrays.}

\author{F. Sgrignuoli}
\affiliation{Department of Electrical and Computer Engineering \& Photonics Center, Boston University, 8 Saint Mary's Street, Boston, Massachusetts 02215, USA}
\author{L. Dal Negro}
\email{dalnegro@bu.edu}
\affiliation{Department of Electrical and Computer Engineering \& Photonics Center, Boston University, 8 Saint Mary's Street, Boston, Massachusetts 02215, USA}
\affiliation{Division of Material Science and Engineering, Boston University, 15 Saint Mary's Street, Brookline, Massachusetts 02446, USA}
\affiliation{Department of Physics, Boston University, 590 Commonwealth Avenue, Boston, Massachusetts 02215, USA}
\begin{abstract}
We investigate light transport in novel three-dimensional scattering systems generated according to subrandom sequences and demonstrate subdiffusive behavior typical of wave transport in disordered systems at the critical point for metal-insulator-transitions but in a wider range of parameters. Specifically, we solve the electromagnetic multiple scattering problem using the Green's matrix spectral theory for aperiodic systems based on Halton, Sobol, and stochastic Latin-Hypercube sequences. By studying the Thouless number and the level spacing statistics of the electromagnetic resonances at different scattering density we demonstrate that light transport in  deterministic Halton and Sobol structures exhibit multifractal behavior characterized by inverse power law scaling of level spacing statistics across a wide range of densities of dipolar scatterers. On the other hand, this scenario is absent in the stochastic Latin-Hypercube array, whose behavior resembles instead standard diffusion in uniform random media. Our findings establish a connection between subdiffusion and subrandom aperiodic order and provide a novel strategy to design three-dimensional structures with multifractal properties over a broad spectral range.
\end{abstract}

\pacs{Valid PACS appear here}
\maketitle
\section{Introduction}\label{Introduction}
The near-field dipole-dipole coupling between randomly located scatterers is considered one of the main reasons preventing the onset of a delocalization-localization-transition (DLT) of vector waves in three-dimensional (3D) disordered systems \cite{SkipetrovPRL,Bellando}. One of the peculiarities of light transport around the DLT is that it acquires subdiffusive and multifractal properties, giving rise to the weak localization regime \cite{Hu,Faez,Evers,Chab,John,Sebbah}. Due to the uncorrelated nature of the uniform disorder model \cite{Sheng,Akkermans,Anderson,DiederikPhotonic}, the DLT is predicted to occur only when a strong magnetic field is applied to a 3D ensemble of two-level atoms \cite{Skipetrov2015,Skipetrov2018,Cottier}. Moreover, uniform random (UR) systems lack efficient design protocols, often limiting their applications to optical design engineering \cite{Balestri,Riboli}. To deal with these problems, novel strategies have been developed to localize electromagnetic fields based on aperiodic order in combination with defects engineering \cite{Freedman} and tailored disorder \cite{Levi}. More recently, it has been shown that aperiodic systems that leverage flat-band physics \cite{WangLoc,SgrignuoliCL} or deterministic aperiodic geometries \cite{Sgrignuoli2019,DalNegroCrystals} can support a delocalization-localization transition as well as multifractality in the optical response across the visible spectrum of aperiodic nanoparticles arrays based on fundamental structures of algebraic number theory \cite{SgrignuoliMF}. 

In the present work, we study light transport in 3D aperiodic systems generated from subrandom sequences and we demonstrate a transition from a diffusive to a subdiffusive, or weak localization, transport regime. Moreover, we found that these novel scattering systems behave similarly to a disordered medium around the critical DLT point, but across a wider range of densities of dipolar particles. Generally, localization arises due to the presence of structural complexity in disorder media \cite{Sheng,Akkermans,Anderson,DiederikPhotonic}. This can result from random fluctuations in the scattering potentials, such as in the case of Anderson localization, or can be introduced deterministically by leveraging aperiodic order, which is the approach that we have chosen in our paper. 

Subrandom sequences fill a $d$-dimensional space more uniformly compared to uncorrelated random ones \cite{Niederreiter,Kuipers,McKay} and they are extensively used in statistical sampling theory where they provide superior accuracy and convergence properties \cite{Lemieux,Morokoff}. Interestingly, we show that the mathematics of subrandom sequences offers unique opportunities for the design of a novel class of complex media with enhanced light-matter interaction properties with respect to standard UR systems. Specifically, using the rigorous dyadic Green's matrix spectral method we study the light transport properties of 3D arrays of electric dipoles geometrically arranged according to the Halton, Sobol, and Latin-Hypercube (LH) subrandom sequences. By performing a scaling analysis of the Thouless number \cite{Thouless} and by evaluating the first-neighbor level spacing statistics of the complex Green's matrix eigenvalues, we observe clear signatures of a transition into a subdiffusive (\emph{i.e.,} weak localization) regime in the Halton and Sobol configurations, which we found to be hyperuniform deterministic structures. In particular, we discover that this transition in subrandom hyperuniform media is described by a level spacing statistics that does not follow the Ginibre's ensemble of random matrix theory and does not exhibit Poisson statistics at large scattering density, as it is for random media in a strong magnetic field \cite{Skipetrov2015,Skipetrov2018,Cottier}. Instead, we find that the probability density function of the level spacing statistics of the Halton and Sobol configurations is well-reproduced by  the Gaussian unitary ensemble (GUE) of random matrices in the diffusion regime and by an inverse power law scaling in the weak localization regime. On the other hand, we find that the structures generated by the Latin-Hypercube stochastic algorithm are not hyperuniform and do not show any signature of weak localization of vector waves. 

By systematically comparing both the structural properties, up to the fourth-level correlation order, and the spectral properties of subrandom media with the ones of uniform random structures we attribute the observed weak localization behavior to the following structural features: (i) a probability density function for the nearest-neighbor separation of the particles that is smaller at short distances from the one of homogeneous Poisson point patterns \cite{Illian}; (ii) the inhibition of long-wavelength density fluctuations. Importantly, these distinctively geometrical characteristics reduce the possible excitation of proximity resonances, which are ``dark" sub-radiant modes localized over just few scatterers \cite{Heller}. In turn, this helps reducing the near-field mediated dipole-dipole interactions, which have been recently identified as key contributions that prevent the occurrence of light localization in 3D uniform random media \cite{Skipetrov2019}. 

\section{Geometrical properties of low discrepancy sequences}\label{SecMethod}
The three-dimensional scattering media considered in this work have been designed using the theory of subrandom sequences. 
This theory is concerned with point sets and sequences having a uniform distribution inside a real interval, such as the distribution of the fractional parts of certain sequences of real numbers $\{x_n\}=x_n-[x_n]$ in the unit interval $I=[0,1)$. Here $[x_n]$ denotes integer part of $x_n$, which is the greatest integer smaller or equal to $x_n$. The fundamental notion is the one of an equidistributed sequence, or a sequence uniformly distributed modulo one, and abbreviated u.d. mod(1). A sequence $x_n$ of real numbers is said to be u.d. mod(1) when the proportion of terms falling inside any half-open sub-interval of $I$ is proportional to the length of that interval. More formally, $x_n$ is said to be u.d. mod(1) is if it satisfies the relation: 
\begin{equation}
\lim_{N\rightarrow\infty} \frac{A([a,b);N)}{N}=b-a
\end{equation}
for every pair of real numbers $a$ and $b$ with $0\leq a< b\leq 1$, where $A([a;b);N)$ denotes the number of terms $x_{n}$ with $1\leq{n\leq{N}}$ for which the fractional parts of $x_{n}$ belong to the interval $[a;b)$ \cite{Kuipers}. Informally, this definition means that a number sequence $x_{n}$ is u.d. mod(1) if every half-open sub-interval of $I$ eventually contains its ``proper share" of fractional parts.

A central theorem in the theory of equidistributed sequences is the Weyl criterion that provides the necessary and sufficient condition for a general  sequence $x_n$ to be u.d. mod(1) in term of the asymptotic behavior of the corresponding exponential sum. The Weyl's theorem \cite{Kuipers}, which can be generalized in any dimension, states that an arbitrary sequence $x_n$ of real numbers is u.d. mod(1) if and only if: 
\begin{equation}\label{Weyl}
\lim_{N\rightarrow\infty}\frac{1}{N}\sum_{n=1}^N e^{2\pi iqx_n}=0
\end{equation}
for all integers $q\neq0$.
We note that the trigonometric sum appearing above coincides with the array factor of kinematic diffraction theory. In particular, its squared modulus is proportional to the far-field diffracted intensity from an array of point scatterers with coordinates $x_{n}$. Therefore, Weyl's theorem implies that large-scale arrays of point scatterers that form a u.d. mod(1) sequence  strongly suppress far-field scattering radiation everywhere, except along the forward direction.

\begin{figure*}[t!]
\centering
\includegraphics[width=0.8\linewidth]{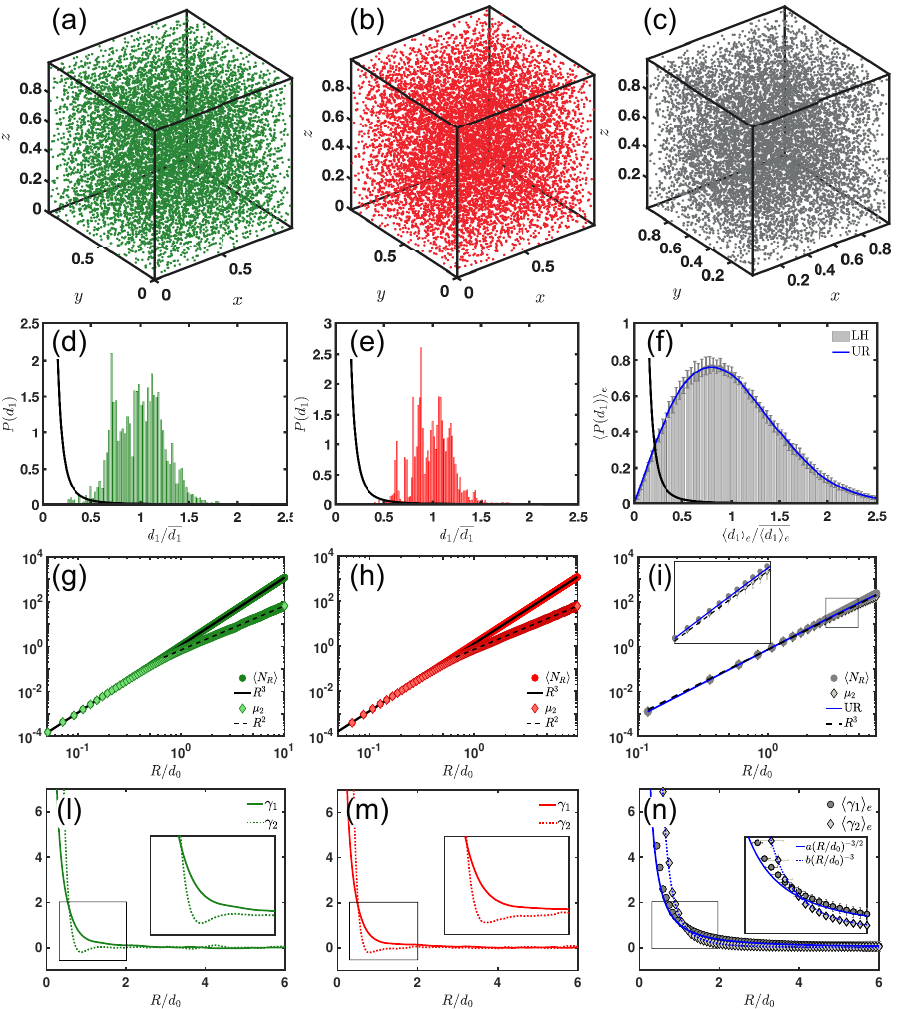}
\caption{(a-c) $10^4$ electric point dipoles spatially distributed by following the Halton (green-points), Sobol (red-points), and latin hypercube (grey-points) subrandom sequences. (d-f) Nearest-neighbor distance probability density function of the point patterns reported on top of each panels. Panel (f) compares the Rayleigh density function, which describes the nearest-neighbor distance distribution of uniform random (UR) point processes \cite{Illian}, with the averaged probability density of the nearest-neighbor
separation of the dipoles in the LH arrays. The average is performed with respect to 1000 different LH realizations. The continuous black lines identify the decay behavior of the dipole-dipole interaction term that it is proportional to $1/r_{ij}^3$. The scaling of the number of variance $\mu_2=\langle N(R)^2\rangle-\langle N(R)\rangle^2$ of $10^4$ particles arranged by following the Halton, Sobol, and LH subrandom sequence within a spherical observation window of radius R is reported in panels (g-i), respectively. Panel (i) displays in blue line the average value of the $\mu_2$ two-point correlation function performed over 2000 different UR realizations. Higher-order correlations functions $\gamma_1$ and $\gamma_2$ of the Halton (l), Sobol (m), and LH (n) arrays. Panel (n) displays  the analytical trends of the skewness and excess of uncorrelated Poisson processes in continuous and dotted blue lines, respectively. The error bars of panels (f), (i), and (n) are the statistical errors evaluated with respect to 2000 different realizations of LH point patterns.}
\label{Fig1}
\end{figure*}

The degree of uniformity of equidistributed sequences is quantified by the mathematical concept of discrepancy. For a one-dimensional sequence $x_{n}$ of $N$ real elements, the discrepancy  $D_N=D_N(x_1,\cdots,x_n)$ is defined by \cite{Kuipers}:
\begin{equation}
D_N=\sup_{0\leq a<b\leq1}\Bigg|\frac{A([a,b);N}{N}-(b-a)\Bigg|
\end{equation}
For any sequence of $N$ numbers, we have: $1/N\leq{D_{N}}\leq{1}$. Clearly, the discrepancy $D_N$ of a sequence $x_n$ will be low if the fraction of points falling into an arbitrary subset of the unit interval is close to be proportional to the length of the interval. An important theorem establishes that a sequence $x_{n}$ is u.d. mod(1) if and only if $\lim_{N\rightarrow\infty}D_N$=0 \cite{Niederreiter}, thus proving a fundamental equivalence between uniform sequences mod(1) and zero-discrepancy sequences. Finite-length sequences with such asymptotic property are often referred to as subrandom, low-discrepancy, or quasirandom sequences. 
They differ substantially from traditional random or pseudorandom sequences, such as the ones utilized in random number generators. In fact, while pseudorandom generators uniformly produce outputs in such a way that each trial has the same probability of falling on equal sub-intervals, subrandom sequences are constrained by the low-discrepancy requirement and each point is generated in a highly correlated manner \cite{Niederreiter} . As a result, subrandom sequences cover a given range of interest more quickly and more evenly than randomly generated numbers (see also Fig\,\ref{Fig1}) \cite{Niederreiter,Kuipers}. For this reason, subrandom sequences are extensively used in statistical modeling techniques, such as the quasi-Monte Carlo method \cite{Lemieux,Morokoff}, where they provide better accuracy and faster numerical convergences \cite{Kuipers,Morokoff,Lemieux,Niederreiter,McKay}. Interestingly, the elements of subrandom sequences can be generated either in a deterministic fashion, as in the case of the Halton and Sobol sequences, or by a stochastic algorithm, as in the case of the Latin Hypercube sequence. 

The principal example of a subrandom sequence is provided by the van der Corput sequence, which represents the fundamental building block for the construction of many others \cite{Kuipers}. It is defined by reversing the base $b$ representation of the number $n$, as explained below. Let $b\geq2$ be a positive integer and $\mathbb{Z}_b=\{0, 1, \cdots, b-1\}$ the least residue system of modulo $b$. Then, every positive integer $n\geq0$ has a unique expansion in base $b$:
\begin{equation}\label{expa}
n=\sum_{k=0}^{m-1} a_k(n)b^k
\end{equation}
where $a_k(n)\in\mathbb{Z}_b$ and $m$ is the smallest integer such that $a_k(n)=0$ for all $j>m$ \cite{Huynh}. To define the van der Corput sequence we have to introduce the so-called ``radical inverse function" \cite{Kuipers}. For an integer $b\geq2$, consider the expansion (\ref{expa}) with $n\in\mathbb{N}$. The function $\phi_b:\mathbb{N}\rightarrow[0,1)$ defined as:
\begin{equation}\label{exphi}
\phi_b(n)=\sum_{k=0}^{m-1}  a_k(n)b^{-k-1}
\end{equation}
is the radical inverse function in base $b$. In reversing the number representation this function makes sure that its values lie in the $(0,1)$ interval. Moreover, $\phi_b(n)$ can be obtained from $n$ by a symmetric reflection of the expansion (\ref{expa}) with respect to the decimal point. The sequence with terms  $x_n=\phi_b(n)$ is the base-$b$ van der Corput sequence, where $b>1$ is a fixed prime number \cite{Kuipers}. This sequence has a discrepancy that scales with the number of elements $N$ as $\sim\ln(N)/N$ \cite{Kuipers}. 

The Halton sequence is a multi-dimensional extension of the van der Corput sequence. To build the Halton sequence we use the van der Corput sequence with different bases for each spatial dimension. In order to generate the 3D Halton array reported in Fig.\ref{Fig1} (a), we have used the sequences $\phi_b(2)$, $\phi_b(3)$, and $\phi_b(5)$ in correspondence to the $x-$, $y-$, and $z-$ coordinates of the electric point dipoles \cite{Kocis}. On the other hand, the generation of the deterministic Sobol configuration shown in Fig.\ref{Fig1} (b) is more sophisticated and requires to permute the order of the elements of the van der Corput sequence. This procedure relies heavily on number theory and on the properties of primitive polynomials to implement permutations along each dimension \cite{Joe}. The theoretical underpinnings regarding the generation of Sobol sequence can be found in Ref.\,\cite{Sobol}. Finally, in Fig.\ref{Fig1} (c) we display a realization of an array generated using the stochastic algorithm known as Latin Hypercube sampling \cite{Huynh}. This method of generating subrandom sequences is fundamentally different from the previous ones since it is no longer deterministic. In its implementation, it divides each dimension of space into $N$ equally probable sections and positions the values of a uniform random variable in each row and in each column of the grid. This step is repeated to distribute random samples in all the sections of the grid with the requirement that there must be only one sample in each row and each column of the grid, ensuring that different random samples are never spaced too closely in each dimension \cite{Huynh}. 

In order to obtain more insights on the structural properties of these novel aperiodic media, we analyzed the probability density function $P(d_1)$ of the nearest-neighbor distance, which is among the most important model utilized in the analysis of spatial point patterns \cite{Illian}. In Fig.\,\ref{Fig1}\,(d-f) we report the calculated statistical distributions of the nearest spacing $d_1$, normalized by the averaged first neighbor spacing $\overline{d_{1}}$. We found that the Halton and the Sobol configurations are characterized by highly fragmented $P(d_1)$ statistics with large amplitude fluctuations, while the distance distribution of the dipole array generated using the stochastic LH algorithm is essentially indistinguishable from the one of a Poisson point process. Indeed, in Fig.\,\ref{Fig1}\,(f) we compare the averaged $\langle P(d_1)\rangle_e$, where the subscript $e$ refers to the ensemble average with respect to 1000 different realizations of the disorder, of an LH array (grey-bars) with the analytical result (blue-line) corresponding to a uniform random (UR) array. For uniform random arrays the nearest-neighbor distance is statistically described by the Rayleigh density function \cite{Illian}:
\begin{equation}
P(d_1)=\frac{d_{1}}{\sigma^2}e^{-d_1^2/2\sigma^2}
\end{equation}
for $d_{1}\geq{0}$ where the variance $\sigma$ is equal to $\sqrt{1/2\pi\mu}$ and $\mu$ is the so-called intensity of the Poisson point process, i.e. the average number of points per unit volume \cite{Illian}. As shown in Fig.\,\ref{Fig1}, the probability to find electric dipoles with a normalized separation lower than 0.5 is very low in the Halton and Sobol arrangements, while the $\langle P(d_1)\rangle_e$ of the LH, which is well described by the Rayleigh distribution, is significantly larger. This distinctive structural difference has a dramatic effect on the strength of the dipole-dipole coupling term, which scales proportionally to $1/r_{ij}^3$ (black-lines in Figs.\,\ref{Fig1}\,(d-f)) as well as on the light localization properties of the arrays. In fact, as it will be shown later in the manuscript, weak light localization occurs in the three-dimensional and deterministic Halton and Sobol arrays but not in the stochastic LH or UR structures.

In order to further characterize the structural properties of subrandom arrays we have evaluated the number variance, skewness, and excess (or kurtosis), which are higher-order correlations functions \cite{Mehta,Bohigas}. Indeed, each of these statistical measures can be defined in terms of the moments 
\begin{equation}\label{moments}
\mu_j=\langle (n-\langle n\rangle)^j\rangle
\end{equation}
where $n$ is the number of elements in an interval of length $L$ and $\langle\cdots\rangle$ represents an average taken over many such intervals throughout the entire system \cite{Mehta,Bohigas}. In particular, the number variance $\mu_2$ is a measure of two-point correlations and enables the identification of the hyperuniform behavior of arbitrary point patterns. Hyperuniformity, a concept introduced by S. Torquato and F. H. Stillinger \cite{Torquato}, is a correlated state of matter characterized by the suppression of long-wavelength density fluctuations \cite{TorquatoReview}. This condition leads to the vanishing of the structure factor $S(\bm{k})\rightarrow0$ in the limit $\bm{k}\rightarrow0$ \cite{TorquatoReview}. Equivalently in 3D structures, hyperuniform systems are characterized by considering the scaling of the fluctuations of the number of points $N_R$ contained within a spherical window of radius R, quantified by the growth of the variance $\mu_2=\langle N_R^2\rangle-\langle N_R\rangle^2$ with respect to $R$. Specifically, a point pattern in $d$ Euclidean dimensions is hyperuniform if $\mu_2$ grows slower than $R^d$. This feature is clearly reported in Figs.\,\ref{Fig1}\,(g) and (h) where the density fluctuations of the Halton (green-diamond markers) and Sobol (red-diamond markers) scale proportionally to $R^2$ (black-dashed line fits), demonstrating their hyperuniform nature. On the other hand, in Fig.\,\ref{Fig1}\,(i) we compare the ensemble averaged (over 1000 stochastic realizations) density fluctuations of the arrays generated according to the LH algorithm (gray-markers) and of a traditional Poisson point process (blue-line). In both cases we show that the number variance grows with the volume of the spherical window instead with surface, \emph{i.e.} $\mu_2\propto R^3$ (black-dashed line fit), indicating that UR and LH structures are not hyperuniform (see also the inset of Fig\,\ref{Fig1}\,(i)\,).  Hyperuniform patterns arise in a variety of biological, mathematical, and physical contexts, which includes glass formations \cite{Chremos}, colloidal packing and hard-sphere packing \cite{Dreyfus,Donev,Zachary}, avian retina \cite{Jiao}, immune systems \cite{Mayer}, large-scale observations of the Universe \cite{Gabrielli}, and in the engineering of novel photonic devices \cite{Florescu,Man,Innocenti,Gorsky}, to cite a few. The present work adds another piece to this puzzle, showing that aperiodic scattering 3D media based on deterministic subrandomness are also hyperuniform. 

Let now discuss the higher-order correlations. The $\gamma_1$ and the $\gamma_2$ functions are defined in terms of the moments (\ref{moments}) as \cite{Mehta,Bohigas}:
\begin{equation}
\gamma_1=\mu_3\mu_2^{-3/2} \,\,\,\,\,\,\,\,\,\,\, \gamma_2=\mu_4\mu_2^{-2}-3 \\ 
\end{equation}
where $\mu_3$ and $\mu_4$ express the 3-level and 4-level correlations, respectively. These high-order correlations functions are reported in Fig.\,\ref{Fig1}\,(l-n) and display, as expected, a very different behavior in the Halton (panel l) and Sobol (panel m) configurations with respect to the LH one (panel n), that was averaged over 2000 different relations. Differently from the uncorrelated Poisson point process (continuous and dotted blue lines in Fig.\,\ref{Fig1}\,(n)), the Halton and Sobol subrandom arrays are characterized by a skewness and an excess that goes to zero in the $R/d_0<1$ range. This difference is attributed to intrinsic higher-order correlation effects \cite{Bohigas}. Moreover, $\gamma_1$ and $\gamma_2$ go to zero for all the arrays when $R/d_0\geq 1$. The approach to zero, however, is faster for the Halton and Sobol arrays demonstrating the effects of third- and fourth-order structural correlations \cite{Bohigas}. On the other hand, Fig.\,\ref{Fig1} (n) shows that the LH array has no structural correlations up to the fourth-level correlation order. Indeed, both their higher-order correlations functions nicely match the analytical expressions $\gamma_1=a(R/d_0)^{-3/2}$ and $\gamma_2=b(R/d_0)^{-3}$ that are valid for an uncorrelated Poisson process. Here the coefficients $a$ and $b$ are equal to $1/2\sqrt{\rho}$ and $1/3\rho$, respectively. The parameter $\rho$ is the scatterers density $N/V$, while $N$ and $V$ are the number of point and the volume, respectively. 

\section{Spectral properties of low discrepancy sequences}\label{Results}
We now investigate the spectral and wave localization properties of 3D electric dipole arrays generated according to Halton, Sobol, and LH subrandom sequences. Multiple scattering effects in these novel scattering media are studied by analyzing the properties of the Green's matrix with elements \cite{RusekPRE,Lagendijk}:
\begin{equation}\label{Green}
G_{ij}=i\left(\delta_{ij}+\tilde{G}_{ij}\right)
\end{equation}
where $\tilde{G}_{ij}$ has the form:
\begin{eqnarray}\label{GreenOur}
\begin{aligned}
\tilde{G}_{ij}=&\frac{3}{2}\left(1-\delta_{ij}\right)\frac{e^{ik_0r_{ij}}}{ik_0r_{ij}}\Biggl\{\Bigl[\bold{U}-\hat{\bold{r}}_{ij}\hat{\bold{r}}_{ij}\Bigr]\\
&- \Bigl(\bold{U}-3\hat{\bold{r}}_{ij}\hat{\bold{r}}_{ij}\Bigr)\left[\frac{1}{(k_0r_{ij})^2}+\frac{1}{ik_0r_{ij}}\right]\Biggr\}
\end{aligned}
\end{eqnarray}
when $i\neq j$ and $0$ for $i=j$. $k_0$ is the wavevector of light, the integer indexes $i, j \in 1,\cdots,N$ refer to different particles, $\textbf{U}$ is the 3$\times$3 identity matrix, $\hat{\bold{r}}_{ij}$ is the unit vector position from the $i$-th and $j$-th scatter while $r_{ij}$ identifies its magnitude. The real and the imaginary part of the complex eigenvalues $\Lambda_n$ ($n\in$ 1, 2, $\cdots$ 3N) of matrix (\ref{Green}) are related to the detuned scattering frequency $(\omega_0-\omega)$ and to the scattering decay $\Gamma_n$ both normalized to resonant width $\Gamma_0$ of a bare dipole \cite{RusekPRE,Lagendijk}. This spectral approach accounts for all the multiple scattering orders of arbitrary arrays of electric scattering point dipoles, so that the multiple scattering process is treated exactly. In addition, this method separates the structural properties of an arbitrary scattering systems from their material characteristics. Therefore, the predictions of the Green's approach should be considered ``universal" in the limit of electric dipole scatterers that is valid for particles with small size parameter $x=k\hat{r}$ ($k$ is the wavenumber and $\hat{r}$ is the particle radius). However, this method can also be extended to include higher-order multipolar resonances \cite{DalNegroElliptic}, which are outside the scope of the present work. The study of the spectral properties of the non-Hermitian Green matrix (\ref{Green}) is an excellent approximation in the case of atom clouds or of metal/dielectric particles whose size is much smaller than the wavelength \cite{Bohren}. Cold atoms might represent a suitable alternative to dielectric materials to experimentally investigate light transport in 3D environments. Indeed, even though state of the art lithographic techniques allowed the realizations of complex three-dimensional polymeric photonic inverted networks \cite{Muller,Nocentini,Zanotto}, the fabrication of \emph{deterministic} volumetric structures embedded in a polymer matrix is one of the key challenges of materials science today. On the other hand, quantum-gas microscopes \cite{Kuhr} enabled the engineering of one \cite{Endres}, two \cite{BarredoScience}, and even three-dimensional \cite{Greiner,WangCold,Nelson,Barredo} optical potentials with arbitrary shape while keeping single-atom control to simulate models from condensed matter physics in high controlled environments. Therefore, novel 3D optical scattering potentials based on engineered subrandom sequences could be effectively achieved \cite{Barredo}, providing suitable platforms to experimentally demonstrate the results of this paper.

To investigate the nature of light localization in these novel 3D aperiodic structures, we have analyzed the scaling of the minimum value of Thouless conductance and the level spacing statistics as a function of the scattering density $\rho/k_0^3$. Here $k_0$ is the vacuum wavenumber. Specifically, we have studied these spectral properties by numerically diagonalizing the 3N$\times$3N Green's matrix (\ref{Green}) that, in the present manuscript, can describe the propagation of light in 3D atomic clouds with subrandom geometries based on the Halton, Sobol, and the LH sequences. 
\begin{figure}[t!]
\centering
\includegraphics[width=\columnwidth]{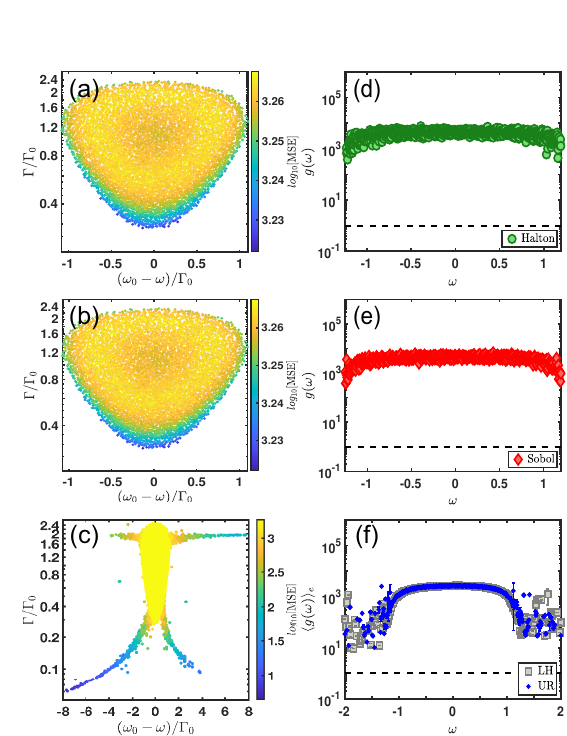}
\caption{Eigenvalues of the electric Green's matrix (\ref{Green}) in the low scattering regime ($\rho/k_0^3=0.001$) of the Halton (a), Sobol (b), and LH (c) subrandom sequence, respectively. Panels (d-f) show the corresponding Thouless numbers as a function of the frequency $\omega$. Panel (f) compares the averaged Thouless number $\langle g(\omega)\rangle_e$ (the subscript $e$ refers to ensemble average with respect to 100 different realizations) of the LH sequence (square-gray markers) with respect to the traditional UR distribution (blue-dots). The error bars are the standard deviations. The dashed-black lines identify the threshold of the diffusion-localization transition.}
\label{Fig2}
\end{figure}

At low scattering density ($\rho/k_0^3=0.001$), all the investigated systems are in the diffusive regime. Their eigenvalue distributions, color coded according to the $\log_{10}$ values of the modal spatial extent (MSE) \cite{SgrignuoliACS}, do not show any particular features. The MSE, which quantifies the spatial extension of any given scattering resonances $\Psi_i$ in the systems, is defined as follows \cite{SgrignuoliACS}:
\begin{equation}
\text{MSE}=\left(\displaystyle\sum\limits_{i=1}^{\hat{N}} \left|\Psi_i\right|^2\right)^2\Big/\displaystyle\sum\limits_{i=1}^{\hat{N}}  \left|\Psi_i\right|^4
\end{equation}
where $\hat{N}$ indicates the total number of scattering resonances. While Figs.\ref{Fig2} (a) and (b) display a very similar distribution of complex scattering resonances delocalized across their 3D geometrical supports (note their very high MSE values), the eigenvalues distributions of the latin hypercube configurations resemble the ones of the standard UR system \cite{SkipetrovPRL, Bellando}. 
These results are corroborated by the behavior of the Thouless number $g$ as a function of the normalized frequency $\omega$ evaluated as \cite{Sgrignuoli2019}:
\begin{equation}\label{Thouless}
g(\omega)=\frac{\overline{\delta\omega}}{\overline{\Delta\omega}}=\frac{(\overline{1/\Im[\Lambda_n]})^{-1}}{\overline{\Re[\Lambda_n]-\Re[\Lambda_{n-1}]}}
\end{equation}
Specifically, we have subdivided the range of the resonance frequencies in 500 equispaced intervals and we have calculated the ratio between the average value of the dimensionless lifetimes and the average spacing of nearest dimensionless resonant frequencies for each frequency sub-intervals. The symbol $\overline{\{\cdots\}}$ in Eq.(\ref{Thouless}) indicates this average operation, while the normalized frequency $\omega$ is the central frequency of each sub-interval used to sample the $\Re[\Lambda_n]$ \cite{Sgrignuoli2019}. As expected, we found that at low scattering density the Thouless number is always larger than one in Fig.\,\ref{Fig2}\,(d-f) for all the analyzed clouds. Moreover, the averaged Thouless number $\langle g(\omega)\rangle_e$ of the LH sequence (square-gray markers) is very similar to the averaged Thouless number of traditional UR systems (blue-dots), as shown in Fig.\,\ref{Fig2}\,(f). The subscript $e$ refers to ensemble average with respect to 100 different Poisson and latin-hypercube different realizations.
\begin{figure}[b!]
\centering
\includegraphics[width=\columnwidth]{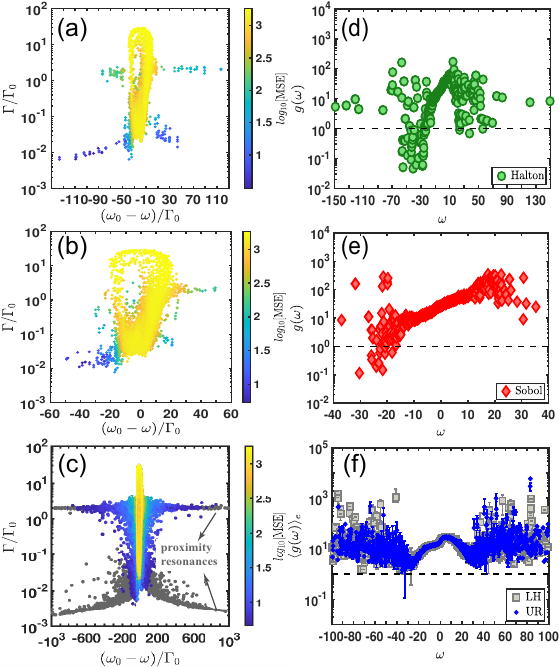}
\caption{Eigenvalues of the electric Green's matrix (\ref{Green}) are shown by points on the complex plane for 2000 electric dipoles arranged by following the Halton (a), Sobol (b), and LH (c) subrandom sequence. These complex eigenvalue distributions are produced when $\rho/k_0^3$ is equal to 0.5. Panels (d-f) show the corresponding Thouless conductances as a function of the frequency $\omega$. Panel (f) compares the averaged Thouless conductance $\langle g(\omega)\rangle_e$ (the subscript $e$ refers to ensemble average with respect to 100 different realizations) of the LH sequence (square-gray markers) with respect to the traditional UR distribution (blue-points). The error bars are the standard deviations. The dashed-black lines identify the threshold of the diffusion-localization transition.}
\label{Fig3}
\end{figure}

On the other hand, at large scattering density $\rho/k_0^3=0.5$ light interacts differently with the two deterministic and hyperuniform 3D subrandom arrays. As shown in Fig.\,\ref{Fig3}, while the stochastic LH subrandom configuration shows a delocalized regime dominated by proximity resonances (dark-grey markers in Fig\,\ref{Fig3}\,(c)), the Halton and the Sobol deterministic configurations are characterized by (i) the formation of spectral gaps, (ii) the absence of proximity resonances, (iii) and a Thouless number lower than one. The absence of sub-radiant dark resonances, attributed to the previously discussed correlation properties of the Halton and Sobol arrays, is reflected in the formation of clear spectral gaps in their  distributions of complex poles. These features reduce drastically the near-field interaction term in eq.\,(\ref{GreenOur}) allowing 3D weak light localization to appear in such systems. On the contrary, this scenario does not occur in traditional UR systems and in the structures based on LH sequences. In these cases, weak localization of vector waves is inhibited due to the strong dipole-dipole interactions resulting from the close particles encounters described by the Rayleigh first-neighbor distance probability distribution, as shown in Fig.\,\ref{Fig1}\,(f).
\begin{figure}[b!]
\centering
\includegraphics[width=\columnwidth]{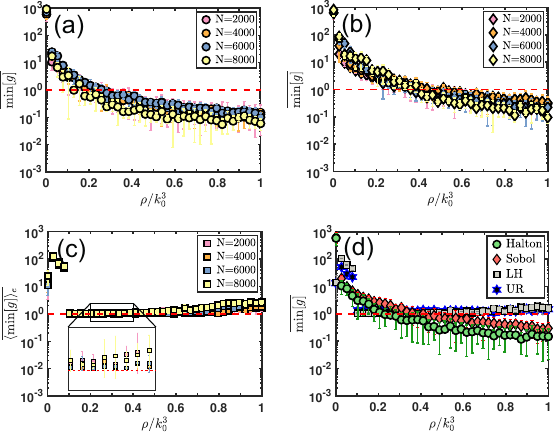}
\caption{Panels (a-c) display the scaling of the minimum value of the Thouless conductance as a function of $\rho/k_0^3$ for Halton (a), Sobol (b), and LH (c) subrandom sequence, respectively. The pastel rose, orange, blue, and yellow markers refer to 2000, 4000, 6000, and 8000 electric dipoles, respectively. The error bars in panel (c) are the standard deviations related to the different disorder realizations. The number of independent configurations were adjusted to ensure a total of at least $2\times10^5$ eigenvalues for each $N$. (d) Minimum of the Thouless conductance as a function of the scattering strength $\rho/k_0^3$ of Halton (green circle-markers), Sobol (red diamond-markers), and LH (grey square-markers) as compared to the uniform random configuration ensemble averaged over 100 different disordered realizations (blue-black pentagrams markers). 100 different realizations were considered also for the LH sequence. The error bars takes into account the different frequency resolutions (see text for more details) used during the Thouless conductance calculations for the Halton and Sobol configurations. In the LH and UR sequences the error bars are statistical errors related to the different disorder realizations. Moreover, the proximity resonances were removed during this analysis.}
\label{Fig4}
\end{figure}

In order to get more insights on the discovered transition from the diffusive to the weak localization regime, we have analyzed the scaling of the minimum value of the Thouless number of arrays with 2000 (pink-markers), 4000 (orange-markers), 6000 (blue-markers), and 8000 (yellow-markers) electric dipoles as a function of the scattering density. The results of this analysis are reported in Fig.\,\ref{Fig4}\,(a), (b), and (c) for Halton, Sobol, and LH clouds, respectively. Specifically, we have evaluated $g=g(\omega)$ by employing eq.\,(\ref{Thouless}) for each $\rho/k_0^3$ value and we have repeated this procedure for different frequency resolutions used in the computation of the Thouless number. Fig.\ref{Fig4} reports their averaged values $\overline{\min[g]}$ and their standard deviations as error bars. In the LH configuration, we performed also an average with respect to different stochastic realizations. The number of independent realizations were adjusted to ensure a total of at least $2\times10^5$ eigenvalues for each N. The scaling of $\overline{\min[g]}$ as a function of the scattering density exhibits a clear trend from $\overline{\min[g]}>$1 to $\overline{\min[g]}<$1 describing a transition into the weak localization regime for the Halton and Sobol structures. Localization begins to take place at $\rho/k_0^3$ approximately equal to 0.25 and 0.4 in the Halton and Sobol configurations, respectively. In contrast, the averaged value of the minimum value of the Thouless number is always larger than one in dipole arrays generated by LH stochastic sequences (see inset of Fig.\,\ref{Fig4}c). Finally, in Fig.\,\ref{Fig4}\,(d) we compare $\overline{\min[g]}$ of the low-discrepancy sequences with the case of traditional random media (pentagram-markers) when $N$=2000. Consistently with the first-neighbor probability distribution, the higher-order correlation analysis, the complex eigenvalues distributions, and the study of the Thouless number, we found that the LH and UR display a very similar behavior. These findings underline a fundamental connection between the structural/geometrical properties of the arrays and their ability/inability to localize electromagnetic waves.

\begin{figure}[t!]
\centering
\includegraphics[width=\linewidth]{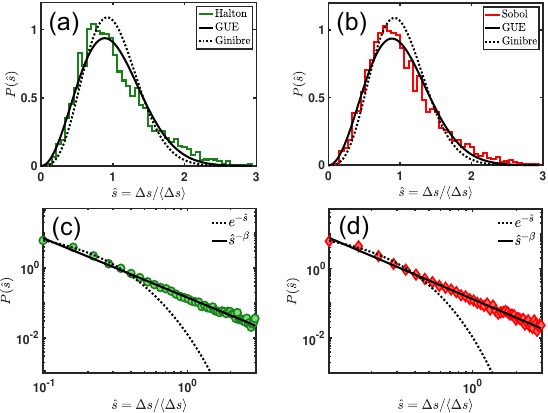}
\caption{Probability distribution functions of level spacing statistic of 6000 Green's matrix eigenvalues for two different scattering regime: $\rho/k_0^3=10^{-3}$ (a-b) and  $\rho/k_0^3=1$ (c-d) for Halton (green) and Sobol (red), respectively. The nearest-neighbor distribution of the eigenvalues of the Gaussian unitary ensemble, described by the eq.\,(\ref{GUE}), and the prediction from the Ginibre's ensemble of random matrices are displayed for comparison in panels (a-b) with continuous and dotted black lines, respectively. Panels (c-d) show that the PDFs of level spacing statistic of the Green's matrix eigenvalues of Halton and Sobol 3D subrandom point patterns does not follow the traditional Poisson distribution $e^{-\hat{s}}$ (black-dotted lines) but a power-law statistic $\hat{s}^{-\beta}$ (black continuous lines) in the large scattering regime. The values of the fitted $\beta$ are 1.7$\pm$0.1 and 1.8$\pm$0.1 in the Halton and Sobol configurations, respectively.}
\label{Fig5}
\end{figure}

In order to further understand the nature of the discovered transition, we have studied the probability density function of the first-neighbor level spacing statistics of the complex Green's matrix eigenvalues $P(\hat{s})$, where $\hat{s}$=$|\Delta\Lambda|/\langle|\Delta\Lambda|\rangle$ is the nearest-neighbor eigenvalue spacing $|\Delta\Lambda|$=$|\Lambda_{n+1}-\Lambda_{n}|$ normalized to the average spacing. This analysis is summarized in Fig.\,\ref{Fig5}. It is well-established that the suppression of the level repulsion phenomenon, \emph{i.e.} $P(\hat{s})\rightarrow0$ when $\hat{s}\rightarrow0$, indicates the transition to localized states in both uniform \cite{Skipetrov2015,Escalante} and non-uniform open-scattering systems \cite{Sgrignuoli2019,DalNegroElliptic}. The distributions of level spacing of Fig.\,\ref{Fig5} show a clear crossover from level repulsion at low optical density (see Fig.\,\ref{Fig5}\,(a-b)) to the absence of level repulsion at large optical density (see Fig.\,\ref{Fig5}\,(c-d)), which is akin to the situation observed in uniform random media under a strong magnetic field \cite{Skipetrov2015}. However, despite this similarity, the observed transition from level repulsion to level clustering presents substantial differences with respect to the UR scenario, indicating that a different localization mechanism governs light interaction in the Halton and Sobol subrandom structures. In fact, we found that at low $\rho/k_0^3$, the distribution of the level spacing
predicted by the Ginibre's ensemble of random matrices, which describes open UR systems \cite{Skipetrov2015,Haake}, does not well-reproduce our spectral statistics, as shown in Fig.\,\ref{Fig5}\,(a,b) in dotted black lines. Instead, an excellent agreement was found using the Gaussian unitary ensemble (GUE) formula \cite{Mehta,Haake}:
\begin{equation}\label{GUE}
P(\hat{s})=\frac{32~\hat{s}^2}{\pi^2}~e^{-4\hat{s}^2/\pi}
\end{equation}
We emphasize that the black curves (continuous and dotted) in Fig.\,\ref{Fig5} (a-b) are not the results of a numerical fitting procedure but are simply obtained using eq.\,(\ref{GUE}) and the expression of the probability density for the normalized complex eigenvalue spacing of the Ginibre's ensemble of random matrices \cite{Skipetrov2015,Haake}. The GUE distribution falls off quadratically for $\hat{s}\rightarrow0$ \cite{Mehta,Haake}, demonstrating that the eigenvalues of the Halton and Sobol exhibit quadratic level repulsion in the low scattering density regime. Interestingly, the GUE distribution (\ref{GUE}) has also been discovered in the spacing of the non-trivial zeros of the Riemann's zeta function \cite{Odlyzko}, whose properties are intimately related to the distribution of prime numbers \cite{Hadamard}. Such a discovery motivated the Montgomery's conjecture \cite{Montgomery} that the pair-correlation function of the non-trivial Riemann's zeros is essentially determined by the properties of random Hermitian matrices. The fundamental connection between the Riemann's zeros and random unitary matrices may provide a fruitful approach to a proof of the Riemann hypothesis \cite{Bombieri}. Interestingly, our findings provide an unexpected connection between the GUE distribution, associated to the distinctive distribution of the Riemann's zeros, and  wave transport (in the low scattering regime) through Sobol and Halton deterministic subrandom structures.  

Additionally, at large optical density we observed a substantial deviation (black-dotted lines in Fig.\,\ref{Fig5} (c-d)) from the Poisson statistics that  typically describes non-interacting exponentially localized energy levels \cite{Mehta,Haake} in UR systems. In contrast, the level spacing distributions for Halton and Sobol configurations, shown by the green-circle and red-diamond markers in Figs.\,\ref{Fig5}\,(c) and (d), are well-reproduced by the inverse power law scaling curves $P(\hat{s})\sim \hat{s}^{-\beta}$ shown by the continuous black lines, with the exponent $\beta$ equal to 1.7$\pm$0.1 and 1.8$\pm$0.1, respectively. In the contest of random matrix theory, it has been demonstrated that this particular distribution is a characteristic of complex systems with multifractal spectra (uncountable sets of hierarchical level clustering) \cite{Cvitanovic,Geisel}. Moreover, this power-law scaling appears to universally describe the transport physics, with values of the exponent $\beta$ in the range $0.5<\beta<2$, of systems exhibiting anomalous diffusion, \emph{i.e.} systems in which the width of a wavepacket $\sigma^2$ increases upon propagation like $t^{2\nu}$ with $\nu\in[0,1]$ \cite{Sokolov,Cvitanovic}. Specifically, such a behavior was observed in one-dimensional scattering systems characterized by incommensurate sinusoidal modulations, in quasi-periodic Fibonacci structures, and in a family of tight-binding Hamiltonians defined on two-dimensional octagonal quasi-periodic tilings \cite{Geisel,Guarneri,Benza}.
The exponents $\beta$ and $\nu$ can be related to the average (box-counting) fractal dimension $D_0$ of the diffusing system through the relation \cite{Cvitanovic,Geisel,DalNegroScirep}
\begin{equation}\label{var}
 \sigma^2(t)\sim t^{2\nu}=t^{2D_0/d}=t^{2(\beta-1)/d}
 \end{equation}
where $d$ is the system dimensionality. By substituting the numbers obtained from the numerical fits of the data in Fig.\,\ref{Fig5} (c-d) into eq.(\ref{var}), we find that the exponent $\nu$ is equal to $0.23\pm0.03$  and $0.27\pm0.07$ for the Halton and Sobol configurations, respectively. The fact that $\nu$ is lower than 0.5 in both configurations indicates that the propagation of wavepackets thought such structures is subdiffusive, potentially enabling novel subdiffusive laser structures that leverage deterministic subrandomness as an effective approach to achieve reduced amplification thresholds and footprints compared to traditional random lasers \cite{Chen}. Interestingly, the behavior that we observed in deterministic subrandom structures closely resembles the electronic transport in 3D weakly disordered systems at the metal-insulator-transition (MIT) where multifractality has been demonstrated \cite{Schreiber} with the subdiffusive exponent $\nu=0.2$ \cite{Schreiber,Schreiber2,Pichard,Sebbah}. This result redefined the standard picture of localization demonstrating that subdiffusion, which is produced by weak localization effects \cite{John}, is an intermediate step between the diffusive and the fully localized regime \cite{Sebbah}. By following this interpretation, the reported crossover between level repulsion and level clustering in Fig.\,\ref{Fig5} can be explained as a transition from a diffusive to a weak localization regime in which the scattering resonances are multifractal and the transport dynamics becomes subdiffusive.

Our findings clearly establish that deterministic subrandom structures strongly reduce dipole-dipole interactions resulting in the reported transition to weak localization of light in 3D complex environments.

\section{Conclusions}

In conclusion, we have systematically investigated the structural and spectral properties of novel electromagnetic wave scattering systems consisting of three dimensional arrays of electric dipoles with subrandom aperiodic order. We demonstrated new types of deterministic aperiodic scattering systems that exhibit in a wide range of parameters the multifractal behavior typical for the critical point of DLT in random media. Specifically, by performing a scaling analysis of the Thouless number and by studying the first-neighbor level spacing statistics of the complex Green's matrix eigenvalues, we have established a clear transition from a diffusive to a weak localization regime in the Halton and Sobol structures, which are characterized by a power-law level spacing at large optical density and by GUE statistics in the diffusive regime.  Moreover, we have also shown that subrandom structures generated by the stochastic  Latin-Hypercube sequence do not show any signatures of light localization. By comparing both the structural, up to the fourth-level correlation order, and the scattering properties of subrandom structures with the ones of uniform random media, we established two properties of primary importance to achieve weak localization of electromagnetic waves: (i) a marked deviation from a Rayleigh probability distribution for the first-neighbor spacing statistics of UR systems, (ii) the suppression of the long-wavelength density fluctuations. These structural features lead to the absence of proximity resonances, and a reduction of the dipole-dipole interactions that should not surpass a critical strength in order to guarantee light localization in 3D. Our analysis clearly shows that the strength of the dipole-dipole coupling between vector dipoles is drastically reduced in the Halton and Sobol configurations due to their structural correlation properties. Recent developments in quantum-gas microscopes have allowed the creation of 3D resonant optical traps of arbitrary shapes while keeping single-atom control. These novel techniques may offer a reliable platform to experimentally demonstrate the results of our manuscript.
\begin{acknowledgments}
This research was sponsored by the Army Research Laboratory and was accomplished under Cooperative Agreement Number W911NF-12-2-0023. The views and conclusions contained in this document are those of the authors and should not be interpreted as representing the official policies, either expressed or implied, of the Army Research Laboratory or the U.S. Government. The U.S. Government is authorized to reproduce and distribute reprints for Government purposes notwithstanding any copyright notation herein.
\end{acknowledgments}
\bibliographystyle{apsrev4-1}
\providecommand{\noopsort}[1]{}\providecommand{\singleletter}[1]{#1}%
%

\end{document}